\documentclass[conference]{IEEEtran}
\usepackage{times}
\IEEEoverridecommandlockouts

\usepackage{cite}

\usepackage{multicol}
\usepackage[bookmarks=true]{hyperref}
\usepackage{xcolor}

\usepackage[final]{graphicx}
\usepackage{bbm}
\usepackage{tikz,pgf}
\usetikzlibrary{shapes}
\usetikzlibrary{patterns}
\usepackage{pgfplots}

\usepackage{float}
\usepackage{amsmath,amssymb}
\usepackage{comment}

\usepackage{etoolbox}

\usepackage{mleftright}
\mleftright

\newtheorem{theorem}{Theorem}

\newtheorem{proposition}{Proposition}

\title{Exploiting Local and Cloud Sensor Fusion in Intermittently Connected Sensor Networks}

\author{\IEEEauthorblockN{Michal Yemini}
\IEEEauthorblockA{Stanford University\\
Stanford, CA, USA\\
Email: michalye@stanford.edu}
\and
\IEEEauthorblockN{Stephanie Gil}
\IEEEauthorblockA{Harvard University\\Cambridge, MA, USA\\
Email: sgil@seas.harvard.edu}
\and
\IEEEauthorblockN{Andrea Goldsmith}
\IEEEauthorblockA{Princeton University\\
Princeton, NJ, USA\\
Email: goldsmith@princeton.edu}
\thanks{The authors gratefully acknowledge support through Intel and through the National Science Foundation (NSF) CAREER award \#1845225. 
}
}

\begin{document}
\maketitle

\begin{abstract}
We consider a detection problem where sensors experience noisy measurements and intermittent communication opportunities to  a centralized fusion center (or cloud). The objective of the problem is to arrive at the correct estimate of event detection in the environment. The sensors may communicate locally with other sensors (local clusters) where they fuse their noisy sensor data to estimate the detection of an event locally.  In addition, each sensor cluster can intermittently communicate to the cloud, where a centralized fusion center fuses estimates from all sensor clusters to make a final determination regarding the occurrence of the event across the deployment area. We refer to this hybrid communication scheme as a \emph{cloud-cluster} architecture. Minimizing the expected loss function of networks where noisy sensors are intermittently connected to the cloud, as in our hybrid communication scheme, has not been investigated to our knowledge. We leverage recently improved concentration inequalities to arrive at an optimized decision rule for each cluster and we analyze the expected detection performance resulting from our hybrid scheme.  Our analysis shows that clustering the sensors provides resilience to noise in the case of low communication probability with the cloud. For larger clusters, a steep improvement in detection performance is possible even for a low communication probability by using our cloud-cluster architecture. 
\end{abstract}


\section{Introduction}

We are at an exciting turning point where the  ubiquity of communications infrastructure enables widespread cloud connectivity, and with it, powerful centralized decision making opportunities. However,  cloud connectivity cannot be guaranteed at all times, particularly for sensors operating over mmWave frequency bands or in complex and potentially remote environments where communication is unreliable. Thus, a new paradigm for centralized decision making that takes intermittent connectivity into account is needed. Currently, decision making in sensor networks typically adopts one of two architectures: i) a centralized architecture that is fully connected, or ii) a distributed architecture, such as peer-to-cloud, where connectivity is intermittent. Adopting either of these extremes can be problematic when the assumption of a continuously connected system is not practical, and, alternatively, requiring fully distributed communication leads to an overly conservative system. A more realistic scenario for multi-sensor systems operating in environments with limited connectivity is that they will have access to a combination of these two communication alternatives, a \emph{hybrid} local \emph{and} cloud network. We call this a \emph{cloud-cluster} communication architecture.  Such hybrid communication architectures give rise to important questions such as 1) \emph{how should the data be fused} at a local level in order to achieve the best global decision making ability at the cloud? and 2) what is the optimal size for the sensor clusters that would provide some \emph{resilience to sensor noise and sporadic connectivity of sensors to the cloud?}  Answering these questions would allow us the necessary insight to best exploit a cloud-cluster communication architecture for multi-sensor decision making.

This paper investigates the best architecture to achieve reliable prediction in the case of multiple sensors detecting an event of interest in the environment.  We employ a hybrid architecture where clusters of sensors pre-process their noisy observations, sending a compressed lower-dimensional aggregate observation to the cloud according to the probabilistic availability of the link. 
We develop a parameterized understanding of the trade-offs involved between architectures; either using larger clusters of sensors approaching a more centralized communication scheme, or, using smaller clusters of sensors approaching a distributed communication scheme.  We show that, depending on the values of important parameters, the cloud-cluster communication architecture may have more resilience to noise and sporadic communication present in real-world environments. These parameters include the individual sensor sensing quality that is quantified by its missed detection and false alarm probabilities, and its probability of accessing the cloud.

There has been much work in the area of determining analytical rules for event detection in clustered sensor networks, among these many works are \cite{Cluster:4102537,Tsitsiklis93decentralizeddetection,Cluster:4407646,Cluster:4608995,Cluster:4957097,Cluster:5751237,Cluster:EURASIP,Cluster:8649751,9095393}. These works consider clustered sensor networks as a network organization scheme to reduce the communication overhead to the fusion center (FC). Sensor networks are often characterized by extreme power and communication constraints and thus the objective in decentralized detection for these systems is to perform well, in their ability to detect an event, while transmitting the smallest number of bits possible. While these works make a significant contribution to our understanding of the clustered sensor networks, they do not consider the sporadic nature of the intermittent connectivity of sensors systems. This aspect of the problem is very important, for example, in mmWave communication systems \cite{5876482,5783993,6732923} 
that are vulnerable to temporary blockages, also known as outages. When a channel is blocked, no information can be passed through it, as its capacity is zero. These blockages occur with positive and non-negligible probability as is modeled in \cite{6834753,7511572,8047278} and they become more frequent as distance between the transmitter and receiver grows. Connectivity is also a common problem  in mobile robotic systems (see \cite{yasamin,pappas,zavlanos,gilIJRR}), where robot location affects both the robot connectivity to the FC, and its event-detection probability.
Minimizing the expected loss function of cloud-cluster sensor networks where sensors are intermittently connected to the cloud was not previously investigated. In this work we show that, using recently improved concentration inequalities, we can approximate the expected loss function caused by detection errors.   We note that like prior works \cite{Cluster:4102537,Tsitsiklis93decentralizeddetection,Cluster:4407646,Cluster:4608995,Cluster:4957097,Cluster:5751237,Cluster:EURASIP,Cluster:8649751}, we do not address the problem of optimizing sensor placement, or how to cluster existing sensors, but rather analyze the performance of system architectures defined over these sensors.

The rest of the paper is organized as follows: Section \ref{sec:system_model} presents the system model and problem formulation. Section \ref{sec:analysis} analyzes  the optimal cloud-cluster decision rules. Section \ref{sec:approx_prob} includes approximation to the optimal decision rules when they are intractable. Section \ref{sec:numerical_results} presents numerical results. Finally, Section \ref{sec:conclusion} concludes the paper.

\section{System Model and Problem Formulation}\label{sec:system_model}
\subsection{System Model}
We consider a team of multiple sensors indexed by $i$, $i\in\{1,\hdots,N\}$, that are deployed to sense the environment and determine if the event of interest has occurred.  We assume that the sensors are noisy, and that their ability to detect the event is captured by the probabilities $P_{\text{MD},s_i}$ of missed detection  and $P_{\text{FA},s_i}$ of false alarm for each agent $i$.  Suppose that there are two hypothesis
$\mathcal H_0$ and $\mathcal H_1$, the first occurs with probability $p_0= 1-p_1$ and the second with probability $p_1$. We denote the random variable that symbolizes the correct hypothesis by $\Xi$, where $\Xi\in\{0,1\}$.  We assume for each agent $i$ that the measured bit $y_i$ may be swapped, due to an error event at the sensor level, with the following probabilities
\begin{flalign*}
 P_{\text{FA},s_i}&= \Pr(y_i=1|\Xi=0),\nonumber\\
P_{\text{MD},s_i} &= \Pr(y_i=0|\Xi=1),
\end{flalign*} 
where $P_{\text{FA},s_i},P_{\text{MD},s_i}\in (0,0.5)$ without loss of generality.   
The sensors have intermittent connectivity to a centralized cloud server, or \emph{FC}. This intermittent connectivity is modeled by a binary random variable $t_i$ that is equal to $1$ if sensor $s_i$ can communicate with the FC and $0$ otherwise. Upon obtaining a communication link to the cloud server, a sensor will transmit sensed information from its cluster of sensors to the cloud.  

In a classical approach,  at the cloud, the FC has the objective of determining whether the event has occurred  after observing the measurements $y_i$ of all communicating sensors. The FC gathers the information it receives from the sensors, 
and aims at estimating the correct hypothesis by minimizing the following expected loss function:
\begin{flalign}\label{eq:error_probability_general2}
E(L) = Pr(\Xi=0)P_{FA}L_{10}+Pr(\Xi=1)P_{MD}L_{01}
\end{flalign}
where $L_{10}$ is the loss caused by false alarm, $L_{01}$ is the loss caused by missed detection, and $P_{FA}$ and $P_{MD}$ are the false alarm and missed detection probabilities at the FC decision, respectively. 
In the classical approach, the FC may suffer from loss of connectivity to many sensors when connectivity is low. On other hand, high connectivity incurs high communication overhead such as scheduling that is undesirable.
To reduce the communication overhead at the FC and also improve network connectivity, we propose an alternative approach to overcome these issues.

We study different communication architectures where the sensors in the system are clustered into teams, and the sensors in each of these teams communicate with one another to arrive at a joint decision. This decision is then forwarded to the FC by a member of the cluster that can communicate with the FC. In this way, a cluster's decision can be forwarded to the FC if at least one sensor in the cluster can communicate with the FC.  Upon receiving the processed measurement from the clusters, the FC estimates the correct hypothesis by minimizing \eqref{eq:error_probability_general2} over all clusters.
We call this hybrid design of the sensor
 communication architecture a cloud-cluster architecture. 
 
\subsection{Problem Formulation}
\begin{figure}
      \centering
      \includegraphics[scale=0.57]{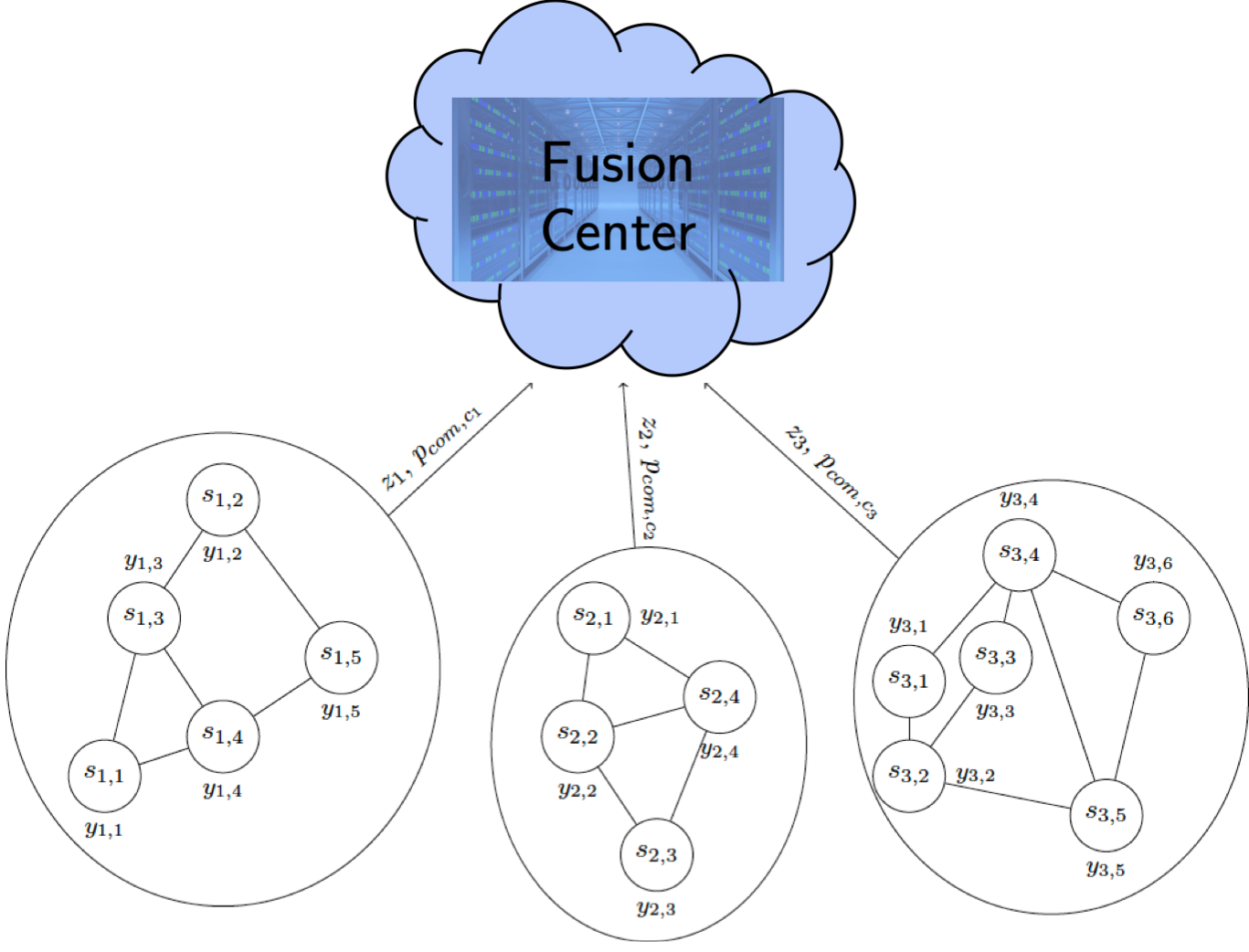}
      \caption{Cloud-cluster architecture.}
      \label{fig:cloud_cluster_arc}
  \end{figure}

We consider a hybrid cloud-cluster system depicted in Fig. \ref{fig:cloud_cluster_arc}. The system is composed of $N_c$ clusters, denoted by $\mathcal{C}_1,\ldots,\mathcal{C}_{N_C}$.
A cluster $\mathcal{C}_j$  communicates with the FC if at least one of the sensors within the cluster
 can communicate with the FC. Let $\tau_j$ be a binary random variable that is equal to one if \textit{cluster} $\mathcal{C}_j$ is communicating with the FC and zero otherwise and  denote   $\boldsymbol{\tau}=(\tau_1,\ldots, \tau_{N_c})$.
  Every sensor cluster $\mathcal{C}_j$ communicating with the cloud sends a pre-processed value $z_j$ that represents the observations of all sensors in cluster $j$. If cluster $\mathcal{C}_j$ cannot communicate with the FC $z_j$ will take an arbitrary deterministic value.
  We denote the vector of the pre-processed values by $\boldsymbol{z}=(z_1,\ldots, z_{N_c})$. The FC at the cloud determines its final decision of whether an event has occurred or not by using the optimal decision rule to minimize~\eqref{eq:error_probability_general2}. 
This optimal decision rule\footnote{We refer the reader to \cite[Chapter 3]{Kay:1993:FSS:151045} for a primer on detection theory and hypothesis testing.} is to choose  hypothesis $\mathcal{H}_1$ if:
\begin{flalign}\label{eq:dec_rule_basic}
\frac{\Pr(\boldsymbol z|H_1,\boldsymbol{\tau})}{\Pr(\boldsymbol z|H_0,\boldsymbol{\tau})} \geq \frac{L_{10}p_0}{L_{01}p_1}
\end{flalign}
and $\mathcal{H}_0$ otherwise. We investigate the following questions: 1) how the data $\boldsymbol{z}$ should be pre-processed at the cluster layer to reduce the expected loss at the FC, 2) how intermittent communication with the cloud impacts performance, and 3) how the overall estimates of missed detection and false alarm  probabilities at the FC are impacted by the communication architecture (i.e., the number of clusters and the number of sensors per cluster).

\section{Analysis}\label{sec:analysis}
This section derives the decision rule and resulting probabilities of missed detection ($P_{MD}$) and false alarm ($P_{FA}$) at the cluster and FC levels for a particular instantiation of our cloud-cluster architecture (i.e., sensor qualities and probabilities of communication). Finding these quantities is made challenging by 1) large cluster sizes (many sensors per cluster) and/or a large number of clusters (many clusters with a small number of sensors per cluster) and 2) the heterogeneous case where individual sensors are allowed to differ in quality, i.e., their individual ability to detect an event, as captured by the probability of missed detection and false alarm \emph{at the sensor level}. Both of these cases result in intractable over $P_{MD}$ and $P_{FA}$ calculations for which we derive an approximation based on concentration inequalities.

\subsection{Cloud-Cluster Communication}
Denote by $s_{j,i}$ sensor $i$ from  cluster $\mathcal{C}_j$. Additionally, denote
 by $p_{\text{com},s_{j,i}}$  the probability that sensor $s_{j,i}$ can communicate with the FC. 
Our cloud-cluster scheme  is aimed at
improving connectivity to FC when the probabilities $p_{\text{com},s_{j,i}}$ are small, and reducing scheduling and communication overheads when the probabilities $p_{\text{com},s_{j,i}}$ are approaching $1$.  
We cluster the sensors into $N_c$ groups.  A cluster of sensors communicates with the FC if one of the sensors comprising the cluster  sees a communication opportunity to the FC. Each cluster estimates the hypothesis and sends its estimation to the FC provided there is a communication opportunity to the FC.

\subsection{Communication probability of clusters}
Let $n_{\mathcal{C}_j}$ be the number of sensors in cluster $\mathcal{C}_j$. 
As we wrote before, a cluster of sensors $\mathcal{C}_j$ communicates with the FC if at least one of the sensors that comprises it can communicate with the FC. It follows that
the probability that the cluster $\mathcal{C}_j$ can communicate with the FC, i.e., $\tau_j=1$, is:
\begin{flalign}
p_{\text{com},\mathcal{C}_j} = 1-\prod_{i=1}^{n_{\mathcal{C}_j}}(1-p_{\text{com},s_{j,i}}).
\end{flalign}
We can see that as we increase the number of sensors in a cluster, i.e., $n_{\mathcal{C}_j}$, the probability of communicating with the cloud, $p_{\text{com},\mathcal{C}_j}$, increases as well. Additionally, as we increase the probability $p_{\text{com},s_{j,i}}$ that a sensor can communicate with the FC, the probability $p_{\text{com},\mathcal{C}_j}$ that the cluster can communicate with the FC is increased as well.

\subsection{Estimations in clusters}

While the objective in the FC is to minimize \eqref{eq:error_probability_general2} directly, the objective in the cluster level is to find the optimal trade-off between the probabilities of false alarm and missed detection.
By the Neyman-Pearson Lemma \cite[Chapter 3]{Kay:1993:FSS:151045} the optimal trade-off can be found by using the likelihood ratio test with a desired threshold $\gamma_j$.  

Let $w_{1,s_{j,i}}=\ln\left(\frac{1-P_{\text{MD},s_{j,i}}}{P_{\text{FA},s_{j,i}}}\right)$ and $w_{0,s_{j,i}}=\ln\left(\frac{1-P_{\text{FA},s_{j,i}}}{P_{\text{MD},s_{j,i}}}\right)$. Denote by
$\mathcal{Y}_{j}$ the sensor measurements in cluster $\mathcal{C}_j$. Following the likelihood ratio test, cluster $\mathcal{C}_j$  chooses the hypothesis $\mathcal{H}_1$ if
\begin{flalign}
\sum_{y_i\in\mathcal{Y}_{j}}\left[w_{1,s_{j,i}}y_i-w_{0,s_{j,i}}(1-y_i)\right]\geq\gamma_j,
\end{flalign} 
and hypothesis $\mathcal{H}_0$ otherwise. 
We note that in case of equality a random decision can be made.
The threshold  $\gamma_j$ is a parameter that the system architecture aims at optimizing to reduce the expected loss at the FC. 

Let $\ell_{\min,j} = -\sum_{i\in\{1,\ldots,n_{\mathcal{C}_j}\}}w_{0,s_{j,i}}$ and $\ell_{\max,j} = \sum_{i\in\{1,\ldots,n_{\mathcal{C}_j}\}}w_{1,s_{j,i}}$.
The threshold $\gamma_j$ can be chosen by searching over the set $\mathcal{L}_j=[\ell_{\min,j},\ell_{\max,j}]$ to minimize \eqref{eq:error_probability_general2}.  
To reduce delay that is caused by recovering at the FC the set of the communicating clusters and sending this information to the clusters, $\gamma_j$ does not depend on the set of communicating clusters. However, our scheme can be adapted to scenarios where the set of communicating clusters changes slowly and so $\gamma_j$ can be optimized for a given realization of communicating clusters.

Now, given the choice of threshold $\gamma_j$ we have that
\begin{flalign}\label{P_errors_inside}
&P_{\text{FA},\mathcal{C}_j} =   \Pr\left(\sum_{y_i\in\mathcal{Y}_{j}}\left[w_{1,s_{j,i}}y_i-w_{0,s_{j,i}}(1-y_i)\right]\geq \gamma_j|\mathcal{H}_0\right), \nonumber\\
&P_{\text{MD},\mathcal{C}_j} = \Pr\left(\sum_{y_i\in\mathcal{Y}_{j}}\left[w_{1,s_{j,i}}y_i-w_{0,s_{j,i}}(1-y_i)\right]<\gamma_j|\mathcal{H}_1\right).
\end{flalign}
 Since the coefficients $w_{1,s_{j,i}}$ and $w_{0,s_{j,i}}$ in \eqref{P_errors_inside} are irrational numbers, the tractable method of calculation presented in \cite{doi:10.1080/00949655.2018.1440294} is not generally applicable to \eqref{P_errors_inside}. Thus we rely on concentration inequalities to approximate the probabilities in \eqref{P_errors_inside} as we describe in Section \ref{sec:approx_prob}.
 
 \subsection{FC Final Decision}
Suppose that the cluster $\mathcal{C}_j$ is communicating with the FC, and denote the data it sends to the FC by $z_i$. 
The optimal decision rule that minimizes  (\ref{eq:error_probability_general2}) is choosing hypothesis $\mathcal{H}_1$ whenever \eqref{eq:dec_rule_basic} holds
and hypothesis $\mathcal{H}_0$ otherwise. 

Let $w_{1,c_{j}}=\ln\left(\frac{1-P_{\text{MD},c_{j}}}{P_{\text{FA},c_{j}}}\right)$ and $w_{0,c_{j}}=\ln\left(\frac{1-P_{\text{FA},c_{j}}}{P_{\text{MD},c_{j}}}\right)$. 
The rule  
\eqref{eq:dec_rule_basic}
can be written as:
\begin{flalign*}
&\sum_{j=1}^{N_c}\tau_j\left[w_{1,c_{j}}z_j-w_{0,c_{j}}(1-z_j)\right]\geq\ln\left(\frac{L_{10}p_0}{L_{01}p_1}\right)=\gamma.
\end{flalign*}

Thus the sensing quality at the FC for a particular realization of the identity of communicating clusters can be written as
\begin{flalign*}
&P_{\text{FA}}(\boldsymbol{\tau}) =  \Pr\left(\sum_{j=1}^{N_c}\tau_j\left[w_{1,c_{j}}z_j-w_{0,c_{j}}(1-z_j)\right]\geq\gamma|\mathcal{H}_0,\boldsymbol{\tau}\right), \nonumber\\
&P_{\text{MD}}(\boldsymbol{\tau}) = \Pr\left(\sum_{j=1}^{N_c}\tau_j\left[w_{1,c_{j}}z_j-w_{0,c_{j}}(1-z_j)\right]<\gamma|\mathcal{H}_1,\boldsymbol{\tau}\right).
\end{flalign*}

The probability of that particular realization of the identity of communicating clusters is
\begin{flalign}
P(\boldsymbol \tau) = \prod_{j=1}^{N_c}p_{\text{com},\mathcal{C}_j}^{\tau_j}(1-p_{\text{com},\mathcal{C}_j})^{1-\tau_j}.
\end{flalign}
This results in the following sensing probabilities
\begin{flalign}\label{eq:error:prob_all_cluster2}
P_{\text{FA}}& = \hspace{-0.3cm}\sum_{\boldsymbol \tau\in\{0,1\}^N}\hspace{-0.2cm}P(\boldsymbol \tau)P_{\text{FA}}(\boldsymbol \tau),\quad
P_{\text{MD}} = \hspace{-0.3cm}\sum_{\boldsymbol \tau\in\{0,1\}^N}\hspace{-0.2cm}P(\boldsymbol \tau)P_{\text{MD}}(\boldsymbol \tau). 
\end{flalign}

\section{Optimizing the Decision Thresholds $\gamma_j$}\label{sec:approx_prob}
Next, we optimize the decision thresholds at the cluster level using the analysis from the previous section.
The complexity of calculating the optimal thresholds $\gamma_j$ is high for the following reasons: first, the optimal thresholds are found by grid search over the sets $\mathcal{L}_1\times\cdots\times\mathcal{L}_{N_c}$.  Additionally, currently no closed form method is known to calculate \eqref{P_errors_inside} and \eqref{eq:error:prob_all_cluster2} efficiently since the coefficient are irrational numbers. Thus, the calculation of these terms is intractable.

\subsection{From grid search to line search}
We overcome the first issue by optimizing each $\gamma_j$ separately using the Gauss-Seidel iterative method. This method optimizes one threshold at a time iteratively until convergence. This approach is considered in relation to sensor network optimization in  \cite{Tsitsiklis93decentralizeddetection}. 
\subsection{Approximating \eqref{P_errors_inside} and \eqref{eq:error:prob_all_cluster2} via concentration inequalities}\label{sec:concentration_clusters_FC}
Now, we explore optimizing the thresholds $\gamma_j$ via concentration inequalities.
We separate the concentration inequalities analysis into two scenarios, both of which are intractable on their own.
Hereafter,  the function $U(\cdot,\cdot,\cdot,\cdot)$ is defined as \eqref{eq:U_def}.

\subsubsection{Large number of sensors in cluster $j$ ($n_{\mathcal{C}_j}\gg1$)} In this case we approximate the false alarm and missed detection probabilities of the decision of cluster $j$ as follows.
\begin{proposition}\label{prop:upper_cluster_error}
Denote 
\begin{flalign*}
\tilde{y}_{j,i} &= w_{1,s_{j,i}}y_{j,i}-w_{0,s_{j,i}}(1-y_{j,i}).
\end{flalign*}
Let
$\alpha_{FA,j}=\gamma_j-\sum_{i=1}^{n_{\mathcal{C}_j}}E(\tilde{y}_{j,i}|\mathcal{H}_0)$ and
$\sigma_{\text{FA},j}^2=\frac{1}{n_{\mathcal{C}_j}}\sum_{i=1}^{n_{\mathcal{C}_j}}\text{var}\left(\tilde{y}_{j,i}-E\left(\tilde{y}_{j,i}|\mathcal{H}_0\right)|\mathcal{H}_0\right)$.
Additionally, let $M_{\text{FA},j} = \max_{i\in\{1,\ldots,n_{\mathcal{C}_j}\}}m_{\text{FA},j,i}$ where
\begin{flalign*}
m_{\text{FA},j,i} = 
&\max\left\{\left\lvert w_{1,s_{j,i}} - E\left(\tilde{y}_{j,i}|\mathcal{H}_0\right)\right\rvert,\left\lvert w_{0,s_{j,i}} + E\left(\tilde{y}_{j,i}|\mathcal{H}_0\right)\right\rvert\right\}.
\end{flalign*}
Then,
\begin{flalign*}
P_{\text{FA},\mathcal{C}_j} \leq U\left(n_{\mathcal{C}_j},\alpha_{\text{FA},j},M_{\text{FA},j},\sigma_{\text{FA},j}^2\right),
\end{flalign*}
whenever $0\leq \gamma_j-\sum_{i=1}^{n_{\mathcal{C}_j}}E(\tilde{y}_{j,i}|\mathcal{H}_0)<n_{\mathcal{C}_j}\cdot M_{\text{FA},j}$.

Additionally, denote
$\alpha_{MD,j}=\sum_{i=1}^{n_{\mathcal{C}_j}}E(\tilde{y}_{j,i}|\mathcal{H}_1)-\gamma_j$ and
$\sigma_{\text{MD},j}^2=\frac{1}{n_{\mathcal{C}_j}}\text{var}\left(E\left(\tilde{y}_{j,i}|\mathcal{H}_1\right)-\tilde{y}_{j,i}|\mathcal{H}_1\right)$.
Let $M_{\text{MD},j} = \max_{i\in\{1,\ldots,n_{\mathcal{C}_j}\}}m_{\text{MD},j,i}$ where
\begin{flalign*}
m_{\text{MD},j,i} = 
&\max\left\{\left\lvert w_{1,s_{j,i}} - E\left(\tilde{y}_{j,i}|\mathcal{H}_1\right)\right\rvert,\left\lvert w_{0,s_{j,i}} + E\left(\tilde{y}_{j,i}|\mathcal{H}_1\right)\right\rvert\right\}.
\end{flalign*}
Then,
\begin{flalign*}
P_{\text{MD},j} &\leq U\left(n_{\mathcal{C}_j},\alpha_{\text{MD},j},M_{\text{MD},j},\sigma_{\text{MD},j}^2\right),
\end{flalign*}
whenever $0\leq \sum_{i=1}^{n_{\mathcal{C}_j}}E(\tilde{y}_{j,i}|\mathcal{H}_1)-\gamma<n_{\mathcal{C}_j}M_{\text{MD},j}$.

\end{proposition}	
We prove Proposition \ref{prop:upper_cluster_error} in Appendix \ref{append:proof_upper_cluster_error}.

\subsubsection{Large number of clusters ($N_{c}\gg 1$)} In this case we approximate the false alarm and missed detection probabilities of the decision of the FC as follows. 

\begin{proposition}\label{prop:upper_FC_error}
Denote, 
\begin{align*}
\tilde{z}_j = \tau_j\left[w_{1,\mathcal{C}_j}z_j-w_{0,\mathcal{C}_j}(1-z_j)\right].
\end{align*}
Let $\alpha_{FA}=\sum_{j=1}^{N_c}E(\tilde{z}_j|\mathcal{H}_1)-\gamma$ and
$\sigma_{\text{FA}}^2=\frac{1}{N_c}\sum_{j=1}^{N_c}\text{var}\left(\tilde{z_j}-E\left(\tilde{z}_j|\mathcal{H}_0\right)|\mathcal{H}_0\right)$.
 Additionally, let $M=M_{\text{FA}} = \max_{j\in\{1,\ldots,N_c\}}m_{\text{FA},j}$ where
\begin{flalign*}
m_{\text{FA},j} = 
\max\left\{\left\lvert w_{1,\mathcal{C}_j} - E\left(\tilde{z}_j|\mathcal{H}_0\right)\right\rvert,\left\lvert w_{0,\mathcal{C}_j} + E\left(\tilde{z}_j|\mathcal{H}_0\right)\right\rvert\right\}.
\end{flalign*}
\end{proposition}	
Then
\begin{flalign*}
P_{\text{FA}} \leq U\left(N_c,\alpha_{\text{FA}},M_{\text{FA}},\sigma_{\text{FA}}^2\right),
\end{flalign*}
whenever $0\leq \gamma-\sum_{i=1}^{N_c}E(\tilde{z}_j|\mathcal{H}_0)<N_c\cdot M_{\text{FA}}$.

Additionally, denote
$\alpha_{MD}=\sum_{j=1}^{N_c}E(\tilde{z}_j|\mathcal{H}_1)-\gamma$
and
$\sigma_{\text{MD}}^2=\frac{1}{N_c}\sum_{j=1}^{N_c}\text{var}\left(E\left(\tilde{z}_j|\mathcal{H}_1\right)-\tilde{z}_j|\mathcal{H}_1\right)$.
Let  $M_{\text{MD}} = \max_{j\in\{1,\ldots,N_c\}}m_{\text{MD},j}$ where
\begin{flalign*}
m_{\text{MD},j} = 
\max&\left\{\left\lvert w_{1,\mathcal{C}_j} - E\left(\tilde{z}_j|\mathcal{H}_1\right)\right\rvert,\left\lvert w_{0,\mathcal{C}_j} + E\left(\tilde{z}_j|\mathcal{H}_1\right)\right\rvert\right\}.
\end{flalign*}
Then
\begin{flalign*}
P_{\text{MD}} &\leq U\left(N_c,\alpha_{\text{MD}},M_{\text{MD}},\sigma_{\text{MD}}^2\right),
\end{flalign*}
whenever $0\leq \sum_{i=j}^{N_c}E(\tilde{z}_j|\mathcal{H}_1)-\gamma<N_cM_{\text{MD}}$.
We prove Proposition \ref{prop:upper_FC_error} in Appendix \ref{append:proof_upper_FC_error}.

Thus we can evaluate the expected loss function and the quality of detection even when exact calculations are intractable. 
\section{Numerical Results}\label{sec:numerical_results}
This section presents numerical results in which we evaluate the performance of the proposed cloud-cluster architecture.
We consider a system with the following characteristics: 500 sensors, to evaluate both the actual and approximate performance, $p(\Xi=1) = 0.4$, $L_{01}=100$ and $L_{10}=150$. Let $\mathcal{U}[a,b]$ denote the uniform distribution on the interval $[a,b]$. To evaluate the performance of the propose approach we compare two systems: a homogeneous one in which $p_{\text{FA},s_i} = 0.2$, $p_{\text{MD},s_i} = 0.3$ for all the sensors in the network, and a heterogeneous system in which for each sensor $i$  we have that  $p_{\text{FA},s_i}\sim \mathcal{U}
[0.16,0.24]$ and $p_{\text{MD},s_i}\sim \mathcal{U}[0.24,0.36]$, that is, both the false alarm and missed detection probabilities of each sensor has a random deviation of 20\% from their values in the homogeneous system. In the heterogeneous setup we average the expected loss of each realization of the false alarm and missed detection probabilities  over 100 realizations. Additionally, in each grid search that we perform for optimizing $\gamma_j$ we use $75$ points per sensor, i.e., a total of $75\times n_{\mathcal{C}_j}$ points.

We use a homogeneous setup with equal cluster size as a tractable setup for which we can calculate the expected loss exactly. We then compare the exact calculation to its approximation.  
In the heterogeneous setup we choose an initial threshold $\gamma_j$ for cluster $\mathcal{C}_j$ by following the same procedure of the homogeneous system assuming that all clusters have the same characteristics as cluster $j$. In the homogeneous setup, for simplicity, we assume that all the thresholds $\gamma_j$ are equal. Additionally,
in both the heterogeneous setup and the approximate calculation in the homogeneous setup we use the approximated missed detection and false alarm probabilities to approximate  $P_{\text{FA},\mathcal{C}_j}$ and  $P_{\text{MD},\mathcal{C}_j}$ presented in Section \ref{sec:concentration_clusters_FC} if $n_{\mathcal{C}_j}>20$. Additionally,  we use the approximated missed detection and false alarm probabilities to approximate  $P_{\text{FA}}$ and  $P_{\text{MD}}$, i.e., the error probabilities at the FC, presented in Section \ref{sec:concentration_clusters_FC} if $N_c>10$. Otherwise we use exact calculations.

\begin{figure}
	\centering
	\includegraphics[scale=0.57,trim={0.5cm 0.5cm 0.5cm 0.6cm},clip]{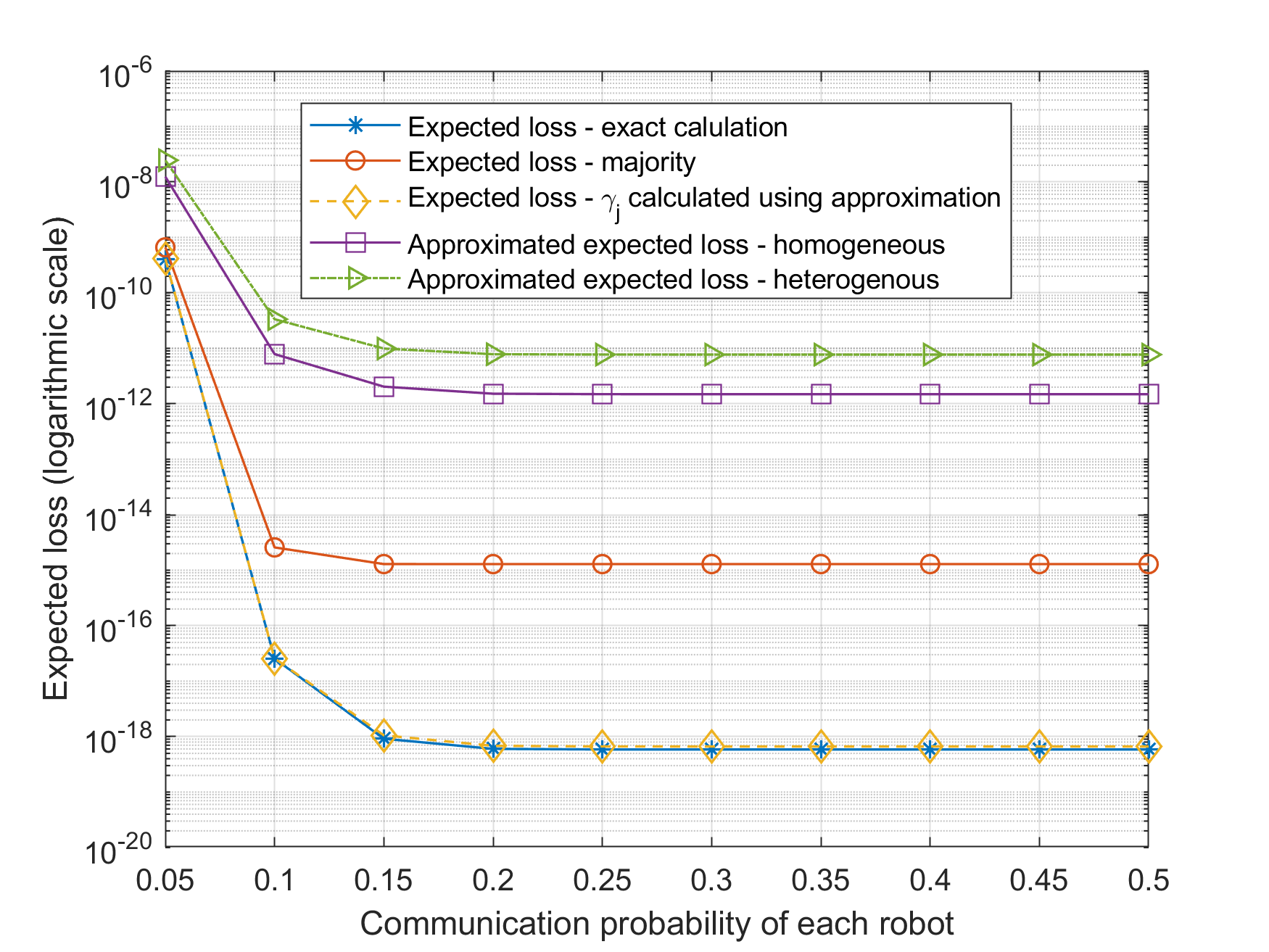}
	\caption{The expected loss function of the communication probability of each sensor for a system with $10$ clusters, each including $50$ sensors. For cloud-cluster architectures we attain a dramatic improvement in performance due to clustering if sensor communication probability to the cloud is at least $0.15$}
	\label{fig_plot_N_c=10}
\end{figure}

\begin{figure}
	\centering
	\includegraphics[scale=0.57,trim={0.5cm 0.2cm 0.5cm 0.6cm},clip]{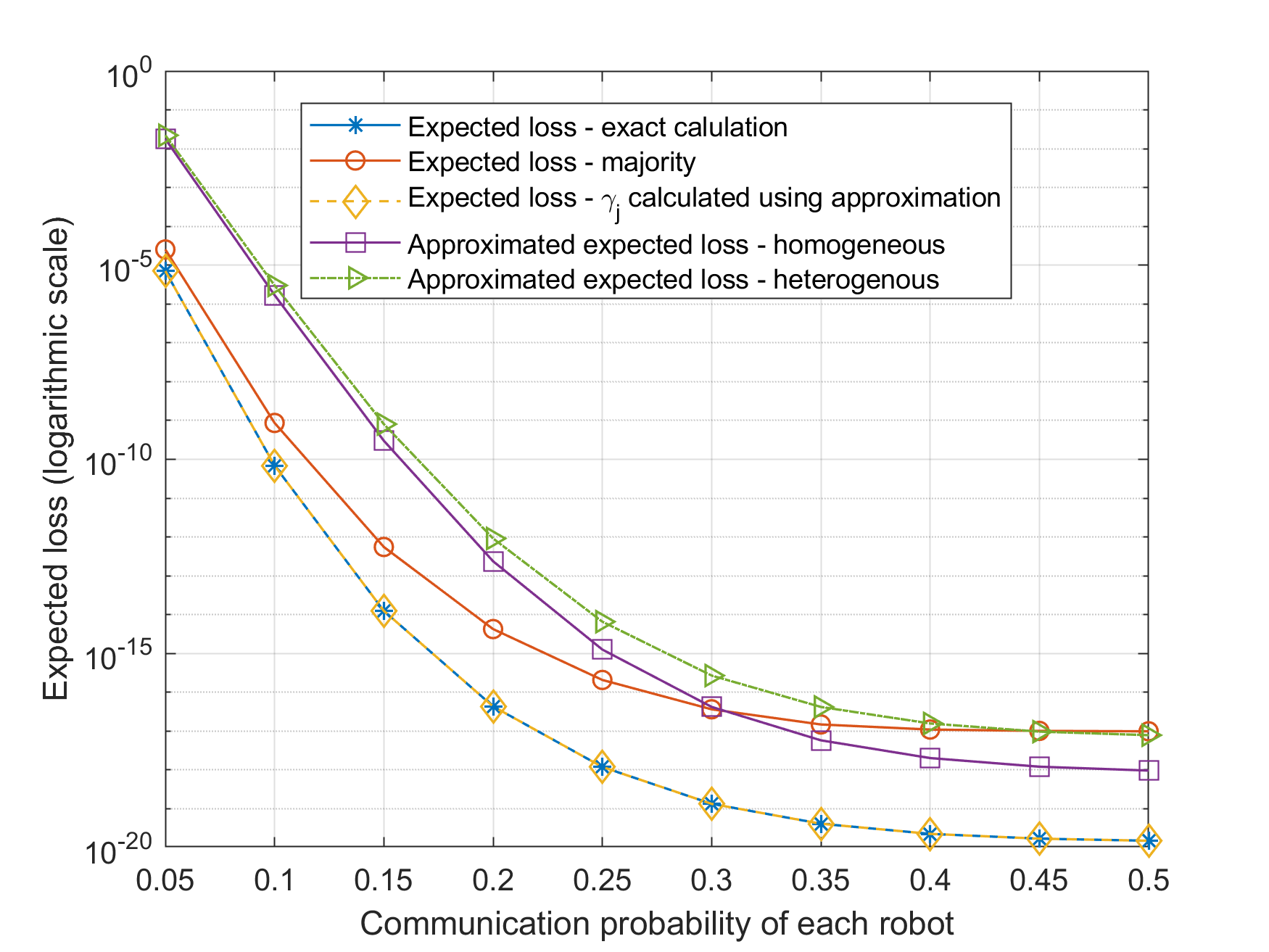}
	\caption{The expected loss function vs. the communication probability of each sensor for a system with $50$ clusters, each including $10$ sensors. For small size clusters, approaching a distributed architecture, higher probability of communication to the cloud is required for better performance.}
	\label{fig_plot_N_c=50}
\end{figure}

\begin{figure}
	\centering
	\includegraphics[scale=0.57,trim={0.5cm 0.2cm 0.5cm 0.6cm},clip]{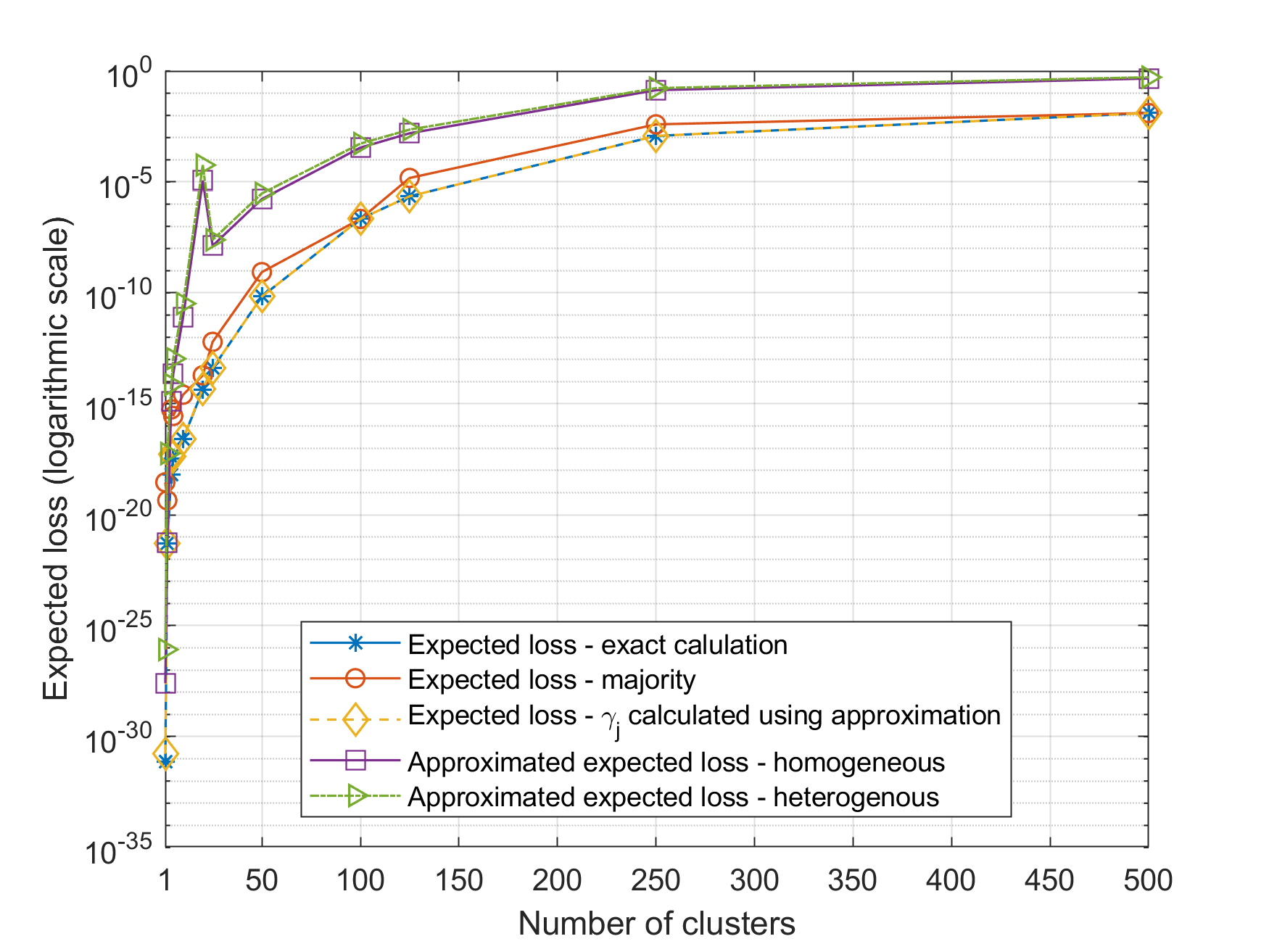}
	\caption{The expected loss function of the number of equal size clusters $N_c$ for $p_{\text{com},s_i}=0.1$. Since connectivity to the FC is low, reducing the number of clusters (more sensors per cluster) increases the chances of communication to the cloud and improves the overall performance.}
	\label{fig_plot_p_c=0.1}
\end{figure}

\begin{figure}
	\centering
	\includegraphics[scale=0.57,trim={0.5cm 0.2cm 0.5cm 0.6cm},clip]{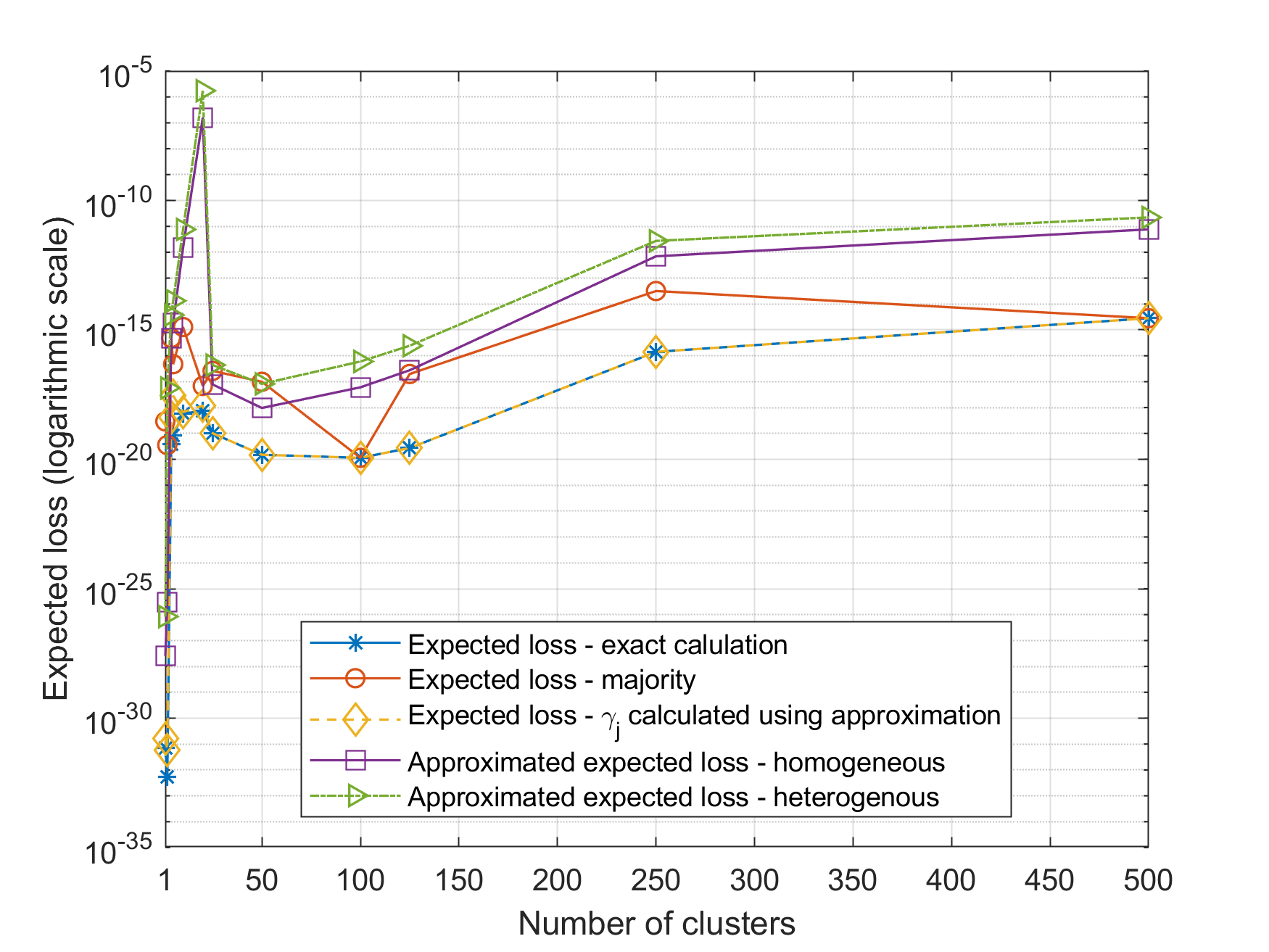}
	\caption{The expected loss function of the number of equal $N_c$ size clusters  for $p_{\text{com},s_i}=0.5$. When connectivity of sensors to the cloud is high, smaller clusters are favored for improving multi-sensor system performance since sensor fusion at the cluster level can be thought of as a form of lossy compression.}
	\label{fig_plot_p_c=0.5}
\end{figure}

Figs. \ref{fig_plot_N_c=10}-\ref{fig_plot_N_c=50} depict the expected loss as a function of the sensor communication probability $p_{\text{com},s}$ for various values of $N_c$ (the number of clusters). 
 Figs. \ref{fig_plot_p_c=0.1}-\ref{fig_plot_p_c=0.5} depict the expected loss as a function  of the number of clusters $N_c$ that comprise the system for various values of sensor communication probabilities $p_{\text{com},s}$. 
Each of the  figures \ref{fig_plot_N_c=10}-\ref{fig_plot_p_c=0.5} includes five lines also denoted in the legends. These are defined as: \\
\textbf{'Expected loss - exact calculation'}: the expected loss of the homogeneous system using exact calculations.\\
\textbf{'Expected loss - majority'}: the expected loss of the homogeneous system in which each cluster makes a majority rule decision where $\gamma_j = \lfloor n_{\mathcal{C}_j}/2\rfloor+1$. The expected loss is calculated exactly.\\
\textbf{'Expected loss - $\gamma_j$ calculated using approximation'}: the exact expected loss that the choice $\gamma_j$ yields, where $\gamma_j$ is optimized using the concentration inequalities depicted in Sec. \ref{sec:concentration_clusters_FC} in the homogeneous scenario. \\
\textbf{'Approximated expected loss - homogeneous'}: the approximate expected loss that is calculated using the concentration inequalities depicted in Sec. \ref{sec:concentration_clusters_FC} in the homogeneous setup.\\
\textbf{'Approximated expected loss - heterogeneous'}: the approximate expected loss that is calculated using the concentration inequalities depicted in Sec. \ref{sec:concentration_clusters_FC} in the heterogeneous setup.

Figs. \ref{fig_plot_N_c=10}-\ref{fig_plot_N_c=50} show that when the number of clusters is large (i.e., each cluster consists of a small number of sensors) the improvement in performance of highly connected systems, compared with that of a sparsely connected systems, is much larger than the contrasting scenario of a system with a small number of clusters. Additionally, Figs. \ref{fig_plot_N_c=10}-\ref{fig_plot_N_c=50} confirm that  optimizing the thresholds $\gamma_j$ using concentration inequalities yields an expected loss that is on par with that of optimizing $\gamma_j$ using exact calculations. Additionally, Figs. \ref{fig_plot_N_c=10}-\ref{fig_plot_N_c=50} depict the gap between the approximate loss function and the exact one for the homogeneous setup and show that our use of the improved Bennet's inequality results in a good approximation for the expected loss function. Finally, while the heterogeneous setup is not tractable, we expect that our use of the improved Bennet's inequality results in a good approximation for the expected loss function for the heterogeneous setup as well. 

Figs. \ref{fig_plot_p_c=0.1}-\ref{fig_plot_p_c=0.5} show that when the communication probabilities of sensors to the FC are low, we observe a monotonic decrease in loss function as we decrease the number of clusters in the exact loss function, this is also observed in the approximated loss function with small deviations when the systems is composed of $4$ clusters.
When the communication probabilities of sensors to the FC are higher, clustering may actually increase the expected loss. This follows because of the single bit compression that happens in the clusters' single bit decisions. That is, there is a trade-off between the error probabilities of the decisions in clusters and that of the FC. Increasing the number of clusters reduces the number of measurements clusters use to make their decisions, and also reduces the communication probability to the FC since clusters include less sensors and thus reduced chances of seeing an opportunity to access the cloud. However, if the communication probability is high, increasing the number of clusters can result in the FC having more measurements to rely on to make
 its final decision.

\section{Conclusion}\label{sec:conclusion}
We consider multi-sensor systems that operate in environments where cloud connectivity is available intermittently. We provide an analytical study of the tradeoffs between different information exchange architectures to support an event detection task. Our results show that if cloud connectivity is reliable, directing sensors to share their sensed values to the cloud for event detection at a centralized fusion center will always perform best. However, in the more likely scenario where cloud connectivity is intermittent, clustering sensors into local neighborhoods where their sensed values are processed and then sent to the cloud during sporadic communication opportunities performs best.  In particular, our results provide insight into the optimal cluster sizes needed to achieve minimum detection loss at the cloud even in the face of noisy sensor data and intermittent communication. Future work can use the results presented here to optimize the locations of sensors such that they attain the recommended cluster sizes for best detection performance over the environment.



\appendices
\section{}
\subsection{Primer on Concentration Inequalities}
We first provide a primer on key concentration inequality results that we will use for the development of our analysis. Since we consider a heterogeneous setup in which the false alarm and missed detection probabilities may vary, we cannot use the concentration inequality \cite{ARRATIA1989125} for the binomial distribution. Instead we use an improved Bennett's  inequality which is known to outperform both Bernstein and Hoeffding's inequalities as well as the Bennet's inequality \cite{doi:10.1080/01621459.1962.10482149}.  

\begin{theorem}[Improved Bennet's inequality \cite{doi:10.1080/03610926.2017.1367818}]\label{theorem:improved_bennett}
	Assume that $x_1\ldots,x_n$ are independent random variables and $E(x_i)=0$, $E(x_i^2)=\sigma_i^2$ and $|x_i|<M$ almost surely.  Additionally, let 
	\begin{align}\label{eq:A_B_def}
	A &= \frac{M^2}{\sigma^2}+\frac{nM}{\alpha}-1,\quad
	B = \frac{nM}{\alpha} -1,
	\end{align}
	and $\Lambda=A-W(Be^A)$, where $W(\cdot)$ is the Lambert $W$ function.
	Let $\sigma^2=\frac{1}{n}\sum_{i=1}^n\sigma_i^2$, then for any $0\leq \alpha<nM$
	\begin{flalign*}
	&\Pr\left(\sum_{i=1}^nx_i\geq \alpha\right) \leq\exp\left[-\frac{\Lambda \alpha}{M}+n\ln\left(1+\frac{\sigma^2}{M^2}\left(e^{\Lambda}-1-\Lambda\right)\right)\right].
	\end{flalign*}
\end{theorem}
Hereafter we use the notation 
\begin{flalign}\label{eq:U_def}
&U(n,\alpha,M,\sigma^2)\triangleq\exp\left[-\frac{\Lambda \alpha}{M}+n\ln\left(1+\frac{\sigma^2}{M^2}\left(e^{\Lambda}-1-\Lambda\right)\right)\right].
\end{flalign}
We note that it is possible to approximate the probability  $\Pr\left(\sum_{i=1}^nx_i\geq \alpha\right)$ using the Gaussian approximation of $\sum_{i=1}^nx_i$. However, the Gaussian approximation may yield smaller approximate probability than the true one, which we want to upper bound. Therefore, for the clarity of presentation, we use the improved Bennet's inequality in our analysis which upper bounds the desired probability in all scenarios. 
\subsection{Proof of Proposition \ref{prop:upper_cluster_error}}\label{append:proof_upper_cluster_error}
Rewrite \eqref{P_errors_inside} as follows
\begin{flalign*}
&P_{\text{FA},\mathcal{C}_j} \hspace{-0.05cm}=\hspace{-0.05cm} \Pr\hspace{-0.05cm}\left[\sum_{i=1}^{n_{\mathcal{C}_j}}\left[\tilde{y}_{j,i}-E\left(\tilde{y}_i|\mathcal{H}_0\right)\right]\geq \gamma_j-\sum_{i=1}^{n_{\mathcal{C}_j}}E\left(\tilde{y}_{j,i}|\mathcal{H}_0\right)|\mathcal{H}_0\right],\nonumber\\
&P_{\text{MD},\mathcal{C}_j}\hspace{-0.05cm}=\hspace{-0.05cm}\Pr\hspace{-0.05cm}\left[\sum_{i=1}^{n_{\mathcal{C}_j}}\left[E\left(\tilde{y}_{j,i}|\mathcal{H}_1\right)-\tilde{y}_{j,i}\right]\hspace{-0.1cm}>\hspace{-0.1cm}\sum_{i=1}^{n_{\mathcal{C}_j}}E\left(\tilde{y}_{j,i}|\mathcal{H}_1\right)-\gamma_j|\mathcal{H}_1\right].
\end{flalign*}
Now, we can use Theorem \ref{theorem:improved_bennett} to upper bound the false alarm probability of the decision of cluster $j$ by substituting
$x_i=\tilde{y_i}-E\left(\tilde{y}_{j,i}|\mathcal{H}_0\right)$, $\alpha=\alpha_{FA,j}=\gamma_j-\sum_{i=1}^{n_{\mathcal{C}_j}}E(\tilde{y}_{j,i}|\mathcal{H}_0)$.
In this case, $\sigma_{i}^2=\sigma_{\text{FA},s_{j,i}}^2=\text{var}\left(\tilde{y}_{j,i}-E\left(\tilde{y}_{j,i}|\mathcal{H}_0\right)|\mathcal{H}_0\right)$ and $M=M_{\text{FA},j} = \max_{i\in\{1,\ldots,n_{\mathcal{C}_j}\}}m_{\text{FA},j,i}$ where
\begin{flalign*}
m_{\text{FA},j,i} = 
&\max\left\{\left\lvert w_{1,s_{j,i}} - E\left(\tilde{y}_{j,i}|\mathcal{H}_0\right)\right\rvert,\left\lvert w_{0,s_{j,i}} + E\left(\tilde{y}_{j,i}|\mathcal{H}_0\right)\right\rvert\right\}.
\end{flalign*}
We denote the resulting constants defined in Theorem \ref{theorem:improved_bennett} by $A_{\text{FA},j}$, $B_{\text{FA},j}$ and $\Lambda_{\text{FA},j}$.  
Thus, by the improved Bennett's inequality we have that
$P_{\text{FA},\mathcal{C}_j} \leq U\left(n_{\mathcal{C}_j},\alpha_{\text{FA},j},M_{\text{FA},j},\sigma_{\text{FA},j}^2\right)$,
whenever $0\leq \gamma_j-\sum_{i=1}^{n_{\mathcal{C}_j}}E(\tilde{y}_{j,i}|\mathcal{H}_0)<n_{\mathcal{C}_j}\cdot M_{\text{FA},j}$.

Similarly, we can use Theorem \ref{theorem:improved_bennett} to upper bound the missed detection probability of cluster $j$ by substituting
$x_i=E\left(\tilde{y}_{j,i}|\mathcal{H}_1\right)-\tilde{y}_{j,i}$,$\alpha=\alpha_{MD,j}=\sum_{i=1}^{n_{\mathcal{C}_j}}E(\tilde{y}_{j,i}|\mathcal{H}_1)-\gamma_j$.
In this case, $\sigma_{i}^2=\sigma_{\text{MD},s_{j,i}}^2=\text{var}\left(E\left(\tilde{y}_{j,i}|\mathcal{H}_1\right)-\tilde{y}_{j,i}|\mathcal{H}_1\right)$ and $M=M_{\text{MD},j} = \max_{i\in\{1,\ldots,n_{\mathcal{C}_j}\}}m_{\text{MD},j,i}$ where
\begin{flalign*}
m_{\text{MD},j,i} = 
&\max\left\{\left\lvert w_{1,s_{j,i}} - E\left(\tilde{y}_{j,i}|\mathcal{H}_1\right)\right\rvert,\left\lvert w_{0,s_{j,i}} + E\left(\tilde{y}_{j,i}|\mathcal{H}_1\right)\right\rvert\right\}.
\end{flalign*}
We denote the resulting constants defined in Theorem \ref{theorem:improved_bennett} by $A_{\text{MD},j}$, $B_{\text{MD},j}$ and $\Lambda_{\text{MD},j}$. 
By the improved Bennet's inequality we have that
$P_{\text{MD},j} \leq U\left(n_{\mathcal{C}_j},\alpha_{\text{MD},j},M_{\text{MD},j},\sigma_{\text{MD},j}^2\right)$,
whenever $0\leq \sum_{i=1}^{n_{\mathcal{C}_j}}E(\tilde{y}_{j,i}|\mathcal{H}_1)-\gamma<n_{\mathcal{C}_j}M_{\text{MD},j}$.

\subsection{Proof of Proposition \ref{prop:upper_FC_error}}\label{append:proof_upper_FC_error}
We  use Theorem \ref{theorem:improved_bennett} to upper bound the false alarm probability of the final decision of the FC by substituting $j$ with $i$ in Theorem \ref{theorem:improved_bennett} and 
$x_j=\tilde{z_j}-E\left(\tilde{z}_j|\mathcal{H}_0\right)$, $\alpha=\alpha_{FA}=\gamma-\sum_{j=1}^{N_c}E(\tilde{z}_j|\mathcal{H}_0)$.
In this case, $\sigma_{j}^2=\sigma_{\text{FA},\mathcal{C}_j}^2=\text{var}\left(\tilde{z_j}-E\left(\tilde{z}_j|\mathcal{H}_0\right)|\mathcal{H}_0\right)$ and $M=M_{\text{FA}} = \max_{j\in\{1,\ldots,N_c\}}m_{\text{FA},j}$ where
\begin{flalign*}
m_{\text{FA},j} = 
\max\left\{\left\lvert w_{1,\mathcal{C}_j} - E\left(\tilde{z}_j|\mathcal{H}_0\right)\right\rvert,\left\lvert w_{0,\mathcal{C}_j} + E\left(\tilde{z}_j|\mathcal{H}_0\right)\right\rvert\right\}.
\end{flalign*}
We denote the resulting constants defined in Theorem \ref{theorem:improved_bennett} by $A_{\text{FA}}$, $B_{\text{FA}}$ and $\Lambda_{\text{FA}}$.  
It follows from the improved Bennett's inequality  that
$P_{\text{FA}} \leq U\left(N_c,\alpha_{\text{FA}},M_{\text{FA}},\sigma_{\text{FA}}^2\right)$,
whenever $0\leq \gamma-\sum_{i=1}^{N_c}E(\tilde{z}_j|\mathcal{H}_0)<N_c\cdot M_{\text{FA}}$.

Similarly, we can use Theorem \ref{theorem:improved_bennett} to upper bound the missed detection probability of the final decision of the FC by substituting $j$ with $i$ in Theorem \ref{theorem:improved_bennett} and 
$x_j=E\left(\tilde{z}_j|\mathcal{H}_1\right)-\tilde{z}_j$, $\alpha=\alpha_{MD}=\sum_{j=1}^{N_c}E(\tilde{z}_j|\mathcal{H}_1)-\gamma$.
In this case, $\sigma_{j}^2=\sigma_{\text{MD},\mathcal{C}_j}^2=\text{var}\left(E\left(\tilde{z}_j|\mathcal{H}_1\right)-\tilde{z}_j|\mathcal{H}_1\right)$ and $M=M_{\text{MD}} = \max_{j\in\{1,\ldots,N_c\}}m_{\text{MD},j}$ where
\begin{flalign*}
m_{\text{MD},j} = 
\max&\left\{\left\lvert w_{1,\mathcal{C}_j} - E\left(\tilde{z}_j|\mathcal{H}_1\right)\right\rvert,\left\lvert w_{0,\mathcal{C}_j} + E\left(\tilde{z}_j|\mathcal{H}_1\right)\right\rvert\right\}.
\end{flalign*}
We denote the resulting constants defined in Theorem \ref{theorem:improved_bennett} by $A_{\text{MD}}$, $B_{\text{MD}}$ and $\Lambda_{\text{MD}}$. 
By the improved Bennet's inequality we have that
$P_{\text{MD}} \leq U\left(N_c,\alpha_{\text{MD}},M_{\text{MD}},\sigma_{\text{MD}}^2\right)$,
whenever $0\leq \sum_{i=j}^{N_c}E(\tilde{z}_j|\mathcal{H}_1)-\gamma<N_cM_{\text{MD}}$.

\bibliographystyle{IEEEtran}

\end{document}